\documentclass[prl,aps,twocolumn,reprint,floatfix,superscriptaddress,longbibliography,amssymb]{revtex4-1}
\usepackage{graphicx}
\usepackage{amsmath,amssymb}
\usepackage{accents}
\usepackage{float}

\newcommand{\be}{\begin{equation}}
\newcommand{\ee}{\end{equation}}
\newcommand{\bea}{\begin{eqnarray}}
\newcommand{\eea}{\end{eqnarray}}
\newcommand{\ben}{\begin{equation*}}
\newcommand{\een}{\end{equation*}}
\newcommand{\ba}{\begin{align}}
\newcommand{\ea}{\end{align}}

\newcommand{\mbf}{\mathbf}
\newcommand{\mrm}{\mathrm}

\newcommand{\sgn}{\mathrm{sgn}}

\begin{document}

\title{Sign-changing photon-mediated atom interactions in multimode cavity QED}

\author{Yudan Guo}
\affiliation{Department of Physics, Stanford University, Stanford, CA 94305}
\affiliation{E.~L.~Ginzton Laboratory, Stanford University, Stanford, CA 94305}
\author{Ronen M. Kroeze}
\affiliation{Department of Physics, Stanford University, Stanford, CA 94305}
\affiliation{E.~L.~Ginzton Laboratory, Stanford University, Stanford, CA 94305}
\author{Varun D. Vaidya}
\affiliation{Department of Physics, Stanford University, Stanford, CA 94305}
\affiliation{E.~L.~Ginzton Laboratory, Stanford University, Stanford, CA 94305}
\affiliation{Department of Applied Physics, Stanford University, Stanford, CA 94305}
\author{Jonathan Keeling}
\affiliation{SUPA, School of Physics and Astronomy, University of St Andrews, St Andrews KY16 9SS UK}
\author{Benjamin L. Lev}
\affiliation{Department of Physics, Stanford University, Stanford, CA 94305}
\affiliation{E.~L.~Ginzton Laboratory, Stanford University, Stanford, CA 94305}
\affiliation{Department of Applied Physics, Stanford University, Stanford, CA 94305}

\date{\today}

\begin{abstract}

Sign-changing interactions constitute a crucial ingredient in the creation of frustrated many-body systems such as spin glasses.  We present here the demonstration of a photon-mediated sign-changing interaction between Bose-Einstein condensed (BEC) atoms in a confocal cavity. The interaction between two atoms is of an unusual, nonlocal form proportional to the cosine of the inner product of the atoms' position vectors. This interaction arises from the differing Gouy phase shifts of the cavity's degenerate modes. Moreover, these Gouy phase anomalies induce an extra pattern of $\mathbb{Z}_2$-symmetry-breaking in the atomic density-wave self-ordering that arises from a nonequilibrium Dicke-type phase transition in the system. This state is detected via the holographic imaging of the cavity's superradiant emission. Together with Ref.~\cite{GouyPRA2018}, we explore this interaction's influence on superradiant phase transitions in multimode cavities. Employing this interaction in cavity QED spin systems may enable the creation of artificial spin glasses and quantum neural networks.

\end{abstract}

\maketitle

The strong atom-photon interactions provided by cavity QED~\cite{Kimble1998} opens new avenues toward exploring quantum many-body physics in a nonequilibrium setting~\cite{Ritsch2013,Sieberer:2016ej,Kirton:2018vv}.  For example, cavity QED with Rydberg atoms provides strong nonlinear interactions between photons~\cite{Peyronel2012} and can lead to topologically nontrivial many-body states~\cite{Schine2016}.  Nonequilibrium Dicke superradiant phase transitions~\cite{Dimer:2007da,Ritsch2013,Kirton:2018vv} and other superradiant transitions~\cite{Black2003,Bohnet2012:steady} have been observed in transversely pumped cavities with thermal atoms~\cite{Zhiqiang2017} and BECs~\cite{Nagy:2010dr,Baumann2010}, including transitions leading to supersolids~\cite{Esslinger2017}, superradiant Mott insulators~\cite{Landig2016,Klinder2015}, and polariton condensates of supermode-density-waves~\cite{Kollar2017} and spinors~\cite{Kroeze:2018wd}.  

Superradiant phase transitions emerge for an ensemble of randomly distributed atoms trapped inside a transversely pumped cavity~\cite{Black2003,Domokos2002}.   Beyond a threshold pump strength, the cavity-photon-mediated interaction energy overcomes the kinetic energy cost associated with the formation of an atomic density wave (DW).  Consequently, the atoms self-organize into a checkerboard pattern on the lattice formed by the transverse pump and cavity mode. The phases of the atomic DW and cavity mode are locked together and  locked to either  $\{0,\pi\}$  with respect to the pump, thus breaking a $\mathbb{Z}_2$ symmetry~\cite{Black2003,Baumann2010,Baumann11}.

In the dispersive limit of cavity QED, where the pump field is not resonant with the cavity modes, the photon field may be adiabatically eliminated. These superradiant phase transitions may then be seen to arise from an effective Hamiltonian with an atom-atom interaction (or spin-spin interaction for spinful atoms) mediated by the exchange of virtually excited cavity photons~\cite{Ritsch2013,Vaidya:2018fp,Kroeze:2018wd}.  Single-mode cavities support infinite-range interactions among the atoms, while multimode cavities provide the means for tuning the range of interactions~\cite{Vaidya:2018fp} and may allow the formation of superfluid liquid crystalline-like states~\cite{Gopalakrishnan2009,Gopalakrishnan2010}.  Photon-mediated interactions might also be possible via the use of photonic waveguides~\cite{Chang:2013bh} and are similar to the phonon-mediated interactions demonstrated among trapped ions~\cite{Porras2004,Kim2009,Britton2012}.

While tunable in range, the interactions among neutral atoms $i,j$ have  been demonstrated with only a fixed-sign coupling $J_{ij}$~\cite{Vaidya:2018fp}.  A wider range of many-body phenomena might be possible if $J_{ij}$ were to flip in sign, because sign-flipping can induce frustrated interactions, as has been demonstrated with ions~\cite{Kim2010}.  With the addition of positional randomness, structural~\cite{Gopalakrishnan2009,Gopalakrishnan2010} and spin glasses~\cite{Gopalakrishnan2011,Strack2011} of atoms in multimode cavities and waveguides~\cite{Douglas:2015hd} may be possible.  These fascinating states exhibit rigidity that arises from a complex---and in some limits, unknown---order and symmetry breaking~\cite{FisherHertz,stein2013spin}.   Creating a tunable-interaction-range spin glass in the quantum-optical setting would provide a novel platform for investigating both how such order emerges, and how quantum phenomena may affect glassy physics. 

\begin{figure*}[t!]
\includegraphics[width = 0.95\textwidth]{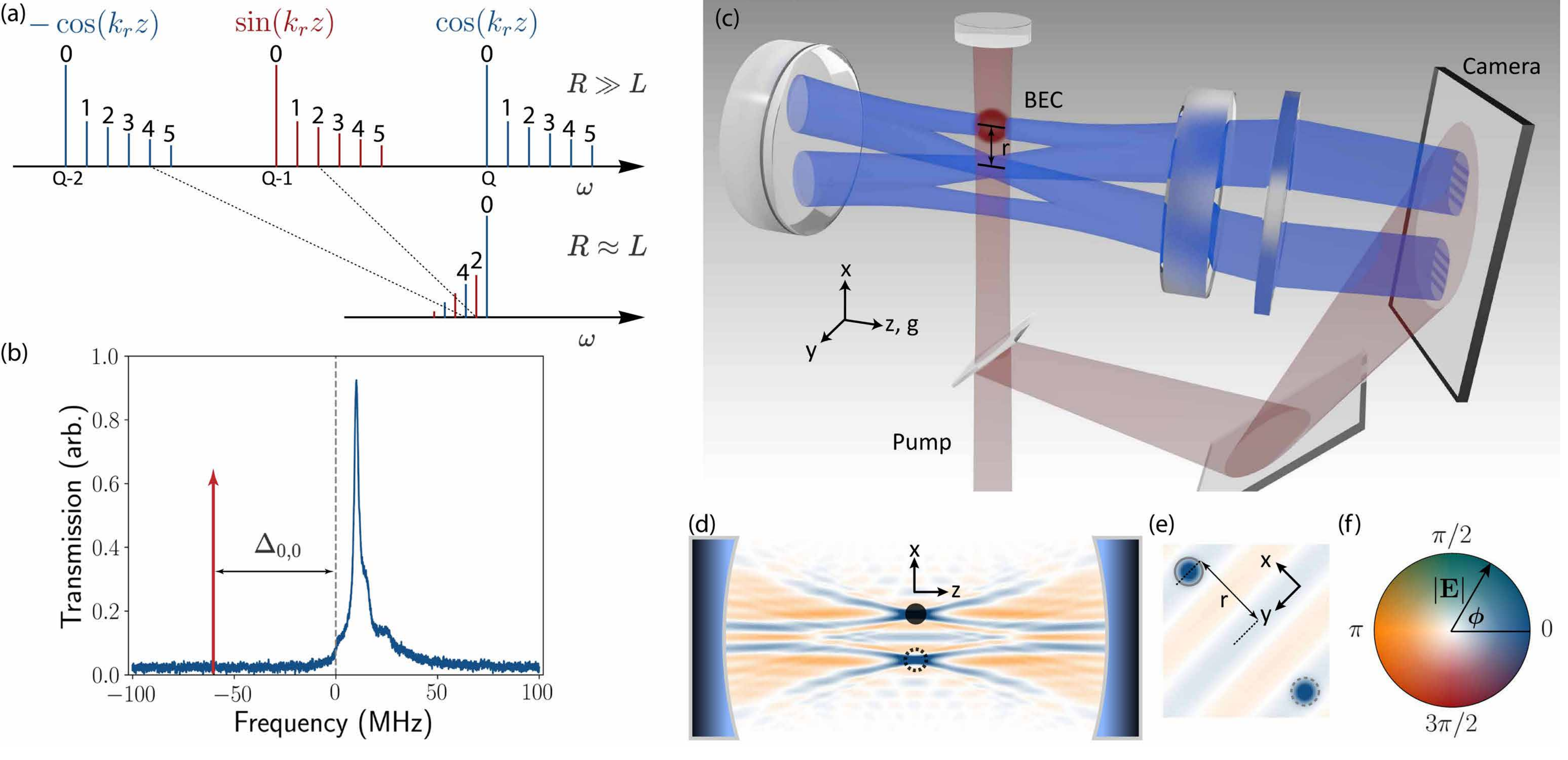}
\caption{(a) Relationship between modes in a near-planar cavity (upper) versus a near-confocal cavity (lower). As $R \rightarrow L$, higher-order transverse modes shift further in frequency than lower-order modes due to differential Gouy phases.  (Near-)degenerate resonances arise at confocality ($R \approx L$) comprised of modes (either even or odd) from different longitudinal families $Q$ with different longitudinal patterns, as indicated. (b) Transmission spectrum of the employed near-confocal cavity. All data below are taken with detuning $\Delta_{0,0} = 60$~MHz, defined from the position of the TEM$_{0,0}$ resonance. (c) Sketch of experimental apparatus showing one of two possible BECs (red sphere) confined within the cavity by optical tweezer traps (not shown). Two images (real and mirror) of the supermode created by the BEC appear in the cavity emission due the fixed parity of the confocal cavity modes~\cite{siegman1986lasers,Kollar2017,Vaidya:2018fp}. Spatial heterodyning of the emitted field is performed by interfering the pump laser (red) and cavity emission (blue) at the EMCCD camera.   (d,e) Simulation illustrating (d) the intracavity field pattern and (e) resulting camera image of the object plane.  Simulated camera image shows two bright spots and an oscillating emission pattern between them.  This oscillating pattern arises from the $U_\text{nonlocal}$ term and has its same functional form.  The bright spots in the far-field cavity emission are generated by the combination of $U^+_\text{local}$ and $U_\text{nonlocal}$.   (f) Color wheel illustrating the complex electric field. }
\label{fig1}
\end{figure*}

In a step toward this goal, we demonstrate here a sign-changing, nonlocal $J_{ij}$ using a multimode cavity.  Previously, we presented a derivation of this term and provided experimental evidence for its existence~\cite{Vaidya:2018fp}.  However, the work neither demonstrated its sign-changing property, nor that it induces a breaking of an additional  $\mathbb{Z}_2$-symmetry coexisting with the checkerboard symmetry described above. The enlarged  $\mathbb{Z}_2\otimes \mathbb{Z}_2$ symmetry-breaking involves the condensation of a DW in either a $\sin{k_rz}$ or $\cos{k_rz}$ pattern along the cavity axis $\hat{z}$ \textit{and} the aforementioned choice of checkerboard state $\delta=\{ 0,\pi\}$ within that pattern---i.e., $\sin{(k_rz+\delta)}$ or $\cos{(k_rz+\delta)}$. Here, $z=0$ is defined at the cavity center,  $k_r = 2\pi/\lambda$, and $\lambda=780$~nm is the cavity and pump wavelength.  
We first discuss  this nonlocal term before presenting results of three experiments. The first and second experiments demonstrate the switching between   $\cos{k_rz}$ or $\sin{k_rz}$ DWs for a cavity with one and two intracavity BECs, resp., while the third demonstrates the sign changing capability of the interaction using two intracavity BECs moved relative to one another. A companion paper~\cite{GouyPRA2018} presents background theory and corroborating experiments in addition to other aspects of interactions induced by Gouy phase anomalies.

The nonlocal interaction term $U_\text{nonlocal}$ arises from  the differing Gouy phase shifts of the degenerate modes of the near-confocal multimode cavity.
Gouy phase anomalies occur in any focused wave and lead to a phase advance as the field propagates through its waist~\cite{Gouy:1890,*Gouy:1891}~\footnote{References~\cite{Boyd:1980ek,siegman1986lasers,Feng*2001,Padgett:2008bi,Visser:2010dx} discuss the many physical interpretations of the Gouy phase anomaly.}.  Fields of higher-order Hermite-Gaussian transverse profiles $\Xi_{l,m}$ exhibit Gouy phase shifts that increase as  $1+l+m$. This causes transverse TEM$_{l,m}$ modes of a cavity with the same longitudinal mode number $Q$ to resonate at different frequencies. However, when special geometrical conditions are met, as, e.g., in a confocal cavity, transverse modes with differing $Q$ become degenerate; see Fig.~\ref{fig1}(a).   At one such degenerate frequency, all modes are either even- or odd-parity.  We employ an even-parity resonance, and therefore, mirror images of the same field amplitude are supported symmetrically across the cavity axis. See Fig.~\ref{fig1}(c).  

The differing Gouy phases of the modes affect the form of the interaction because the photon-mediated interaction in a multimode cavity arises from the exchange of photons in a superposition of all available modes at the positions of the two atoms~\cite{Gopalakrishnan2009,Gopalakrishnan2010,Vaidya:2018fp}. When accounted for in the sum over all modes, the Gouy phases contribute an additional interaction energy $U_\text{nonlocal}$ to the local interaction. The form of the nonlocal term is derived in Refs.~\cite{Vaidya:2018fp,GouyPRA2018} to be $U_\text{nonlocal}(\mbf{r}_i,\mbf{r}_j)= J_0\mathcal{D}_\text{nonlocal}(\mbf{r}_i,\mbf{r}_j)\cos{k_r x_i}\cos{k_r x_j}$, where $\mbf{r}_i$ are $(x,y)$ coordinates of atom $i$, $\mathcal{D}_\text{nonlocal}(\mbf{r}_i,\mbf{r}_j)= \cos{(2\mbf{r}_i\cdot\mbf{r}_j/w_0^2)}/4\pi$, and $w_0=35$~$\mu$m is the TEM$_{0,0}$ mode waist~\footnote{The mode waist $w_0$ is defined as the $e^{-2}$ intensity. Note Ref.~\cite{Vaidya:2018fp} uses $e^{-1}$ intensity, which removes the factor of 2 from the cosine argument in $U_\text{nonlocal}$.}.  The coupling strength is $J_0 =g_0^2\Omega^2/\Delta_a^2\Delta_{0,0}$, where $g_0 = 2 \pi {\times} 1.47(3)$~MHz is the vacuum Rabi rate for an atom coupled to the center of the TEM$_{0,0}$ mode, $\Omega^2$ is proportional to the pump intensity, and $\Delta_a = -2 \pi {\times} 102$~GHz is the detuning of the pump from the atomic excited state. The position-dependent prefactors $\cos{k_r x_i}$ appearing in the interaction arise due to the standing-wave pump~\footnote{Reference~\cite{Vaidya:2018fp} derived the nonlocal term under conditions of a traveling-wave pump.  This work and Ref.~\cite{GouyPRA2018} consider a standing-wave pump because this likely to be used in the future to implement Ising spin models.}. The local interaction terms are comprised of the real and mirror image terms $U^\pm_\text{local}(\mbf{r}_i,\mbf{r}_j) = U_\text{local} (\mbf{r}_i,\mbf{r}_j) \pm U_\text{local} (\mbf{r}_i,-\mbf{r}_j)$~\cite{Vaidya:2018fp,GouyPRA2018}, where the $\pm$ correspond to  even (odd) resonances; we employ even. 

In addition to $U_\text{nonlocal}$, the Gouy phases induce a division of the cavity resonances into two classes with alternating out-of-phase longitudinal DW patterns; see Fig.~\ref{fig1}(a).  At an even-mode confocal cavity resonance, the total mode function is $\Phi_{Q,l,m}(x,y,z)\propto \Xi_{l,m}(x,y)\cos{k_rz}$ for $l+m~\mrm{mod}~4 =0$ modes, while $\Phi_{Q,l,m}(x,y,z)\propto \Xi_{l,m}(x,y)\sin{k_rz}$ for  $l+m~\mrm{mod}~4 = 2$ modes~\footnote{This is true close to the center of the cavity and when we fix the longitudinal pattern of the $\mrm{TEM}_{00}$ employed to be $\cos{k_r z}$~\cite{GouyPRA2018}.}. Thus, while in a single-mode cavity $H\propto J_0\cos{k_r z_i}\cos{k_r z_j}$, in a confocal cavity, the total interaction is
\be
U\propto U_c(\mbf{r}_i,\mbf{r}_j)\cos{k_r z_i}\cos{k_r z_j} + U_s(\mbf{r}_i,\mbf{r}_j)\sin{k_r z_i}\sin{k_r z_j},\nonumber
\ee
where $U_{c,s} = U^+_\text{local}\pm U_\text{nonlocal}$~\cite{Vaidya:2018fp}.  Moreover,  while the atomic wavefunction may be expanded as $\Psi = \psi_0 + \sqrt{2}\psi_c  \cos{k_r x}\cos{k_r z}$ in a single-mode cavity, an additional atomic field is required in a confocal cavity: $\Psi = \psi_0 + \sqrt{2} \cos{k_rx}[\psi_c  \cos{k_r z} + \psi_s \sin{k_r z}]$.
Here, $\psi_{c,s}$ are the wavefunctions describing the fraction of atoms organized into the orthogonal sine versus cosine quadratures of the longitudinal profile; $\psi_0$ is the initial BEC wavefunction in the optical dipole trap~\footnote{This expansion is valid under the experimentally satisfied condition of low momentum excitation.}. The $\mathbb{Z}_2\otimes\mathbb{Z}_2$ order parameters associated with the transition are the fractions of atoms acquiring a $\lambda$-periodic density modulation in either of the two DWs patterns and the $\delta$ phase of the wave therein; in terms of these wavefunctions, the order parameters are $\chi_{c,s} = (\psi_0 \psi^{*}_{c,s} + \psi^{*}_0 \psi_{c,s})/N$, where $N$ is the BEC population. Each $\chi$ may be viewed as a pseudospin with max/min value $\pm1$; the sign of $\chi$ indicates the relative pseudospin alignment. For  BECs at $\mbf{r}_i$ and $\mbf{r}_j$, one may transform the system's light-matter interaction into an effective spin interaction Hamiltonian of the form   $H_{ij} = -J_{ij}(\chi_{ci}\chi_{cj}-\chi_{si}\chi_{sj})$ after spatial integration~\cite{Supp}.  Here,  $J_{ij}\propto NJ_0 \mathcal{D}_\text{nonlocal}(\mbf{r}_i,\mbf{r}_j)$ and $N$ is each BEC's population. The total effective single-BEC Hamiltonian interaction is $H_1 =  H_{ii}$.  The BEC  organizes into  $\chi_c$ or $\chi_s$ depending on which DW pattern minimizes $U_{ii}$, i.e., whether $J_{ii}$ is positive or negative.  Likewise, for two BECs of equal size and shape, $H_2 =  H_{ii} + H_{jj} + 2H_{ij}$.

\begin{figure}[t!]
\includegraphics[width=0.49\textwidth]{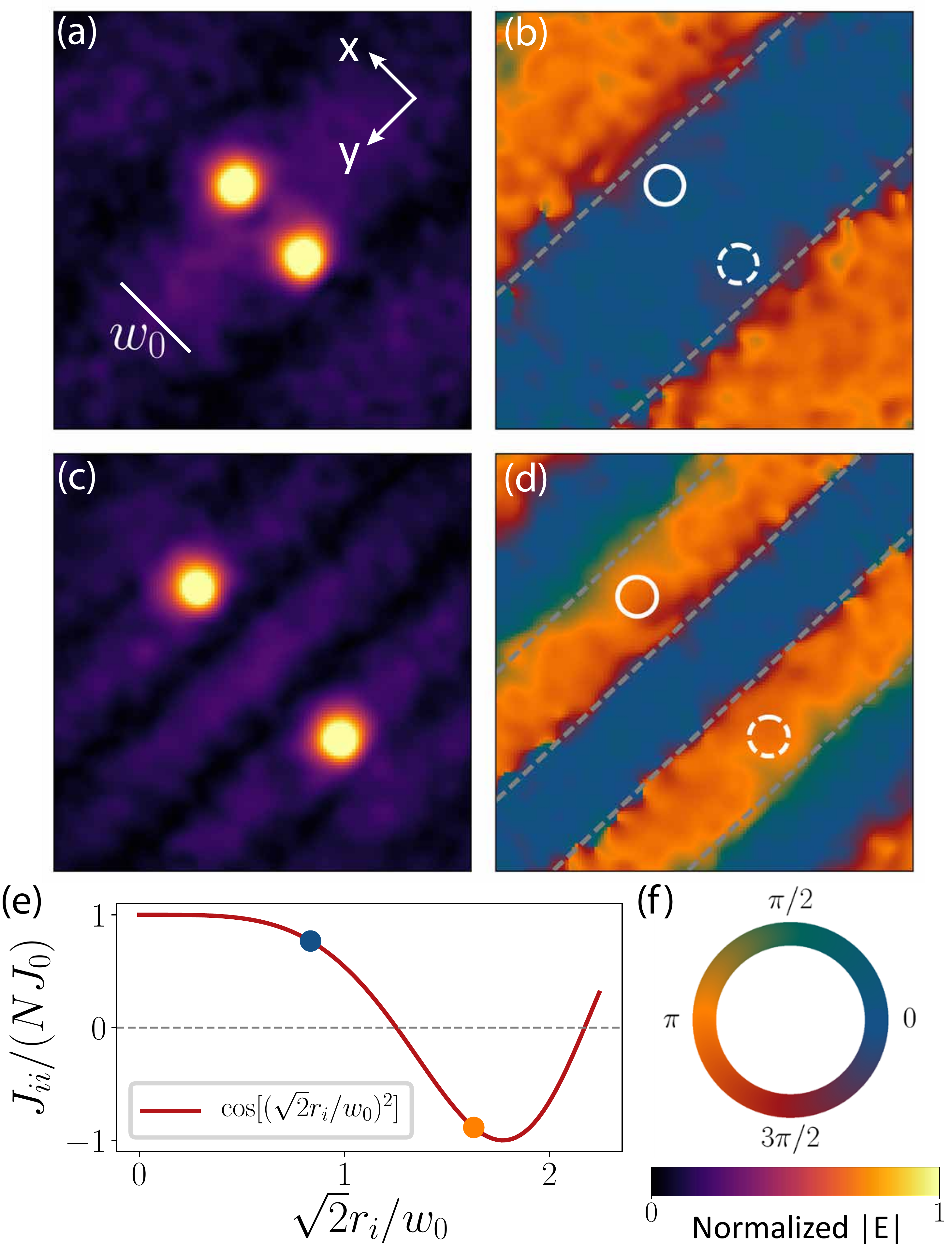}
\caption{(a--d) Extracted superradiant field at the two different positions marked in (e). (a,c) Plots of the extracted normalized field amplitude. The weak modulation arising from the $U_\text{nonlocal}$ atom-photon coupling can be seen as a cosine emission pattern of the crossed bow-tie portion of the cavity field; see Fig.~\ref{fig1}(c--e).  Brighter emission due to $U_\text{local}$ is localized around the BEC (and its mirror image). (b,d) Plots of the corresponding phase data. The dotted lines mark the location of the nodes in the cosine $U_\text{nonlocal}$ pattern as determined from a functional fit to $U_\text{nonlocal}$. The phase of the electric field flips by $\pi$ (while the periodicity shortens) as the BEC's position $\mbf{r}$ is moved across a node in the cosine pattern.   (e) Plot of the functional form of  $J_{ii}$. The blue  and orange dots mark the position of the BEC for the superradiant emission images above.  The observed  phase change is consistent with the flipped sign of $J_{ii}$. (f) Color scale for extracted phase and electric field amplitude, where the phase at $r=0$ is set to 0.}
\label{fig2}
\end{figure}

The experimental apparatus is shown in Fig.~\ref{fig1}(c). The BECs contain  ${\sim}2 {\times} 10^5$ $^{87}$Rb atoms  in the $|F=1,m_F=-1 \rangle$ state.  Optical tweezers position and confine each BEC in a tight trap of diameter $<$10~$\mu$m---smaller than $w_0$. See Refs.~\cite{Kollar2015,Vaidya:2018fp,Supp} for BEC preparation and optical tweezing procedures. To measure the field amplitude and  phase of the superradiant emission, the cavity field and  part of the pump are interfered on an EMCCD camera.  This spatial heterodyne measurement is holographically reconstructed to provide the cavity field amplitude and phase; see Fig.~\ref{fig1}(c-f) and Refs.~\cite{Kroeze:2018wd,Schine:2018ui}. 

Cavity field-emission measurements may be interpreted as cavity-enhanced Bragg scattering: in the organized phase, the transverse pump light is Bragg scattered into the cavity mode from the atomic checkerboard pattern.  The phase of the coherently scattered light is therefore directly correlated with the phase of the  DW.  In addition, in a near-confocal cavity, organization into  $\chi_{c}$ ($\chi_{s}$)  is heralded by a 0 ($\pi$) phase shift between the cavity emission from the position of the BEC and its mirror image versus that from the bow-tie interference pattern between them~\cite{GouyPRA2018}.  This phase shift may be traced back to the $\pm$-sign difference between the $U_\text{local}$ and $U_\text{nonlocal}$ terms in $U_{c,s}$ \cite{GouyPRA2018}.  Figure~\ref{fig2} presents observations of this effect, where the amplitude and phase of superradiant emission from a single BEC at two different positions $\mbf{r}_i$ is shown.  These data demonstrate the ability to tune the DW order from a cosine to sine pattern by controlling $\mbf{r}_i$.  However, the additional $\mathbb{Z}_2$ phase $\delta$ of the checkerboard DW cannot be independently measured with a spatial heterodyne measurement using a single BEC~\footnote{This phase difference can be detected with a temporal heterodyne measurement~\cite{Black2003,Baumann2010,Kollar2017}.}.

\begin{figure}[t!]
\includegraphics[width = 0.49\textwidth]{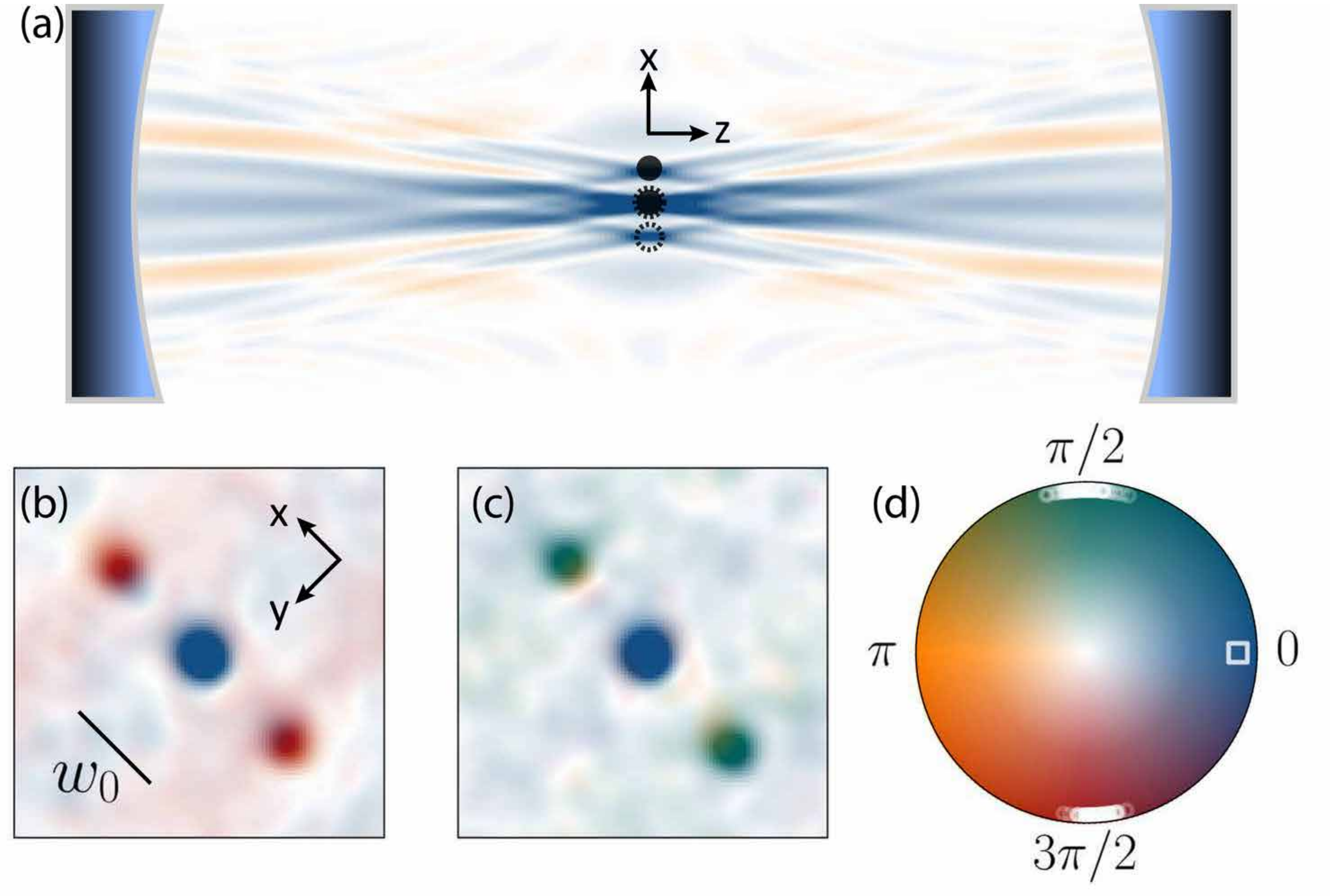}
\caption{(a) Simulated intracavity field for the fixed arrangement of two BECs, as discussed in the main text. The black circles mark the position of the two BECs. (b,c) The measured electric field for two different realizations of the experiment.  The $\pm\pi/2$ phase difference between the two BECs indicates that the BEC at $r=0$ is in a cosine DW, while the other is in a sine wave DW. The sign-flips are indicative of the relative $\mathbb{Z}_2$-symmetry-breaking of the checkerboard patterns within the two DWs.  (d) Color disk for the plotted electric field. The white circular markers register the phase difference between the two spots in 186 shots of the experiments. We measured 92 shots of $\pi/2$ and 94 shots $3\pi/2$. The square marker indicates the reference phase of the $r=0$ BEC: the phase of the light at $r=0$ is set to 0 since we choose cosine DWs to scatter light with 0 relative phase.}
\label{fig3}
\end{figure}

\begin{figure}[t!]
\includegraphics[width = 0.49\textwidth]{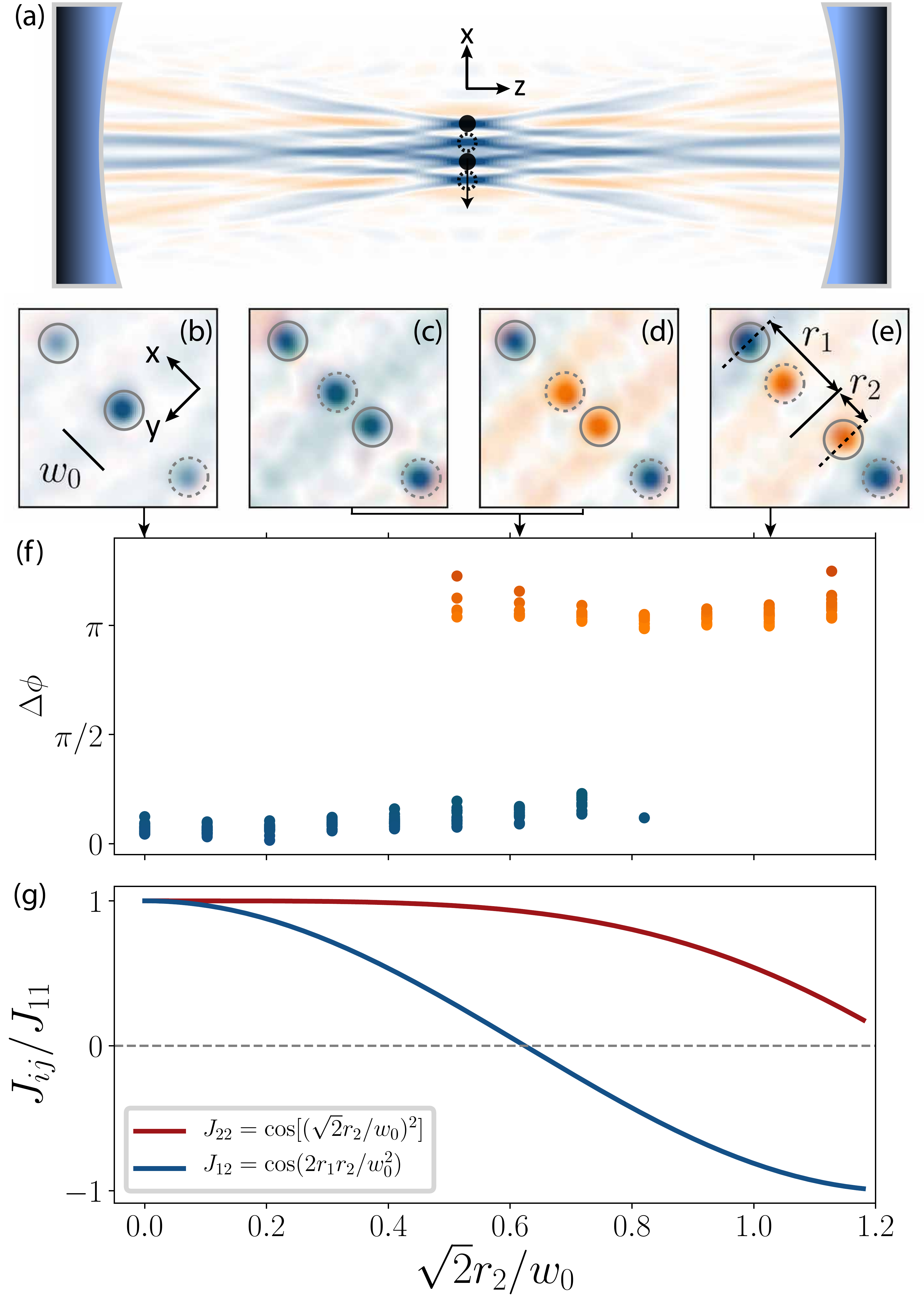}
\caption{(a) Simulated intracavity field for two BECs on either side of the cavity center.  $r_1$ of the first BEC is set so that $J_{11}/N=1$.  The black arrow indicates the direction along which the second BEC is moved a distance ${r}_2$ from center. (b--e) Examples of measured  fields versus ${r}_2$. Panels (c,d) taken at the $r_2$ where $J_{12}=0$.  Random $\pi$ phase flips are observed at this position and in the vicinity of small $J_{12}$.  (f) Phase difference between the BECs' spots versus $r_2$. Twenty data points are plotted for each $r_2$.  (g) Calculation of the  self-interaction $J_{22}$ and cross-interaction $J_{12}$ versus ${r}_2$. Positive $J_{22}$ (and $J_{11}$) ensures that $\cos{k_rz}$ is energetically  favorable  for both BECs until the sign flip in $J_{12}$ causes the second BEC to  condense into opposite pseudospin alignment with $\sgn \{\chi_{c1} \chi_{c2}\} = -1$ .  See Fig.~\ref{fig3}d for color disk scheme. Similar to Fig.~\ref{fig3}, the phase of the light at $r_1$ is set to 0.}
\label{fig4}
\end{figure}

Measurements of $\chi_{c,s}$ are possible using two intracavity BECs.   Detection of the BECs'  sine versus cosine DW pattern  and their relative pseudospin checkerboard state  is possible since a $\chi_c$ DW is $\pm\pi/2$ out of phase from a $\chi_s$ DW, where the $\pm$-sign reflects the relative checkerboard state.  That is, the $\sgn \{\chi_{c,s}\} = +1$  DW is $\delta=\pi$ out of phase from the $\sgn \{\chi_{c,s} \} = -1$ DW. To observe this effect, we place one BEC at $r_1=0$ and the other at $r_2 = \sqrt{\pi} w_0/\sqrt{2}$ along $\hat{x}$, as shown in Fig.~\ref{fig3}a. This sets  $J_{11} = -J_{22} = N$ and the cross-term $U_{12}=0$ because the $J_{ii}$ terms cause the two BECs to prefer different DW quadratures.  That is, the cross-terms in $U_{12}$ vanish $\chi_{c1}\chi_{c2} = \chi_{s1}\chi_{s2} = 0$ since $\chi_{c1} \neq 0$ \& $\chi_{s1} = 0 $ for the first BEC and $\chi_{c2} = 0$ \& $\chi_{s2} \neq 0 $ for the second.  This is shown in the measured electric fields of Fig.~\ref{fig3}(b,c).  We see that the phase of light emitted at the $r\neq0$ BEC (and along the bow-tie interference fringe at which it is located) is indeed shifted by $\pm\pi/2$. Because each BEC is free to choose between the $\mathbb{Z}_2$-symmetric checkerboard states  within the preferred DW profile, we observe a random, nearly 50/50  distribution in relative sign over the course of multiple experimental realizations. This lack of $\mathbb{Z}_2$-broken-symmetry bias indicates the absence of inter-BEC coupling (i.e., $U_{12}=0$), as intended~\footnote{Finite-size effects likely bias the symmetry-breaking a small amount~\cite{Baumann11}.}.

Having demonstrated the DW-pattern-shifting effect of $U_\text{nonlocal}$, we now present observations of its sign-changing character.  Again, we use two BECs, but fix one BEC at $r_1 = \sqrt{2 \pi} w_0$, which sets $J_{11}= N$, while moving the position $r_2$ of the second BEC in a range of positions satisfying $J_{22}>0$; see Fig.~\ref{fig4}.  This causes each BEC to energetically prefer organization into the same pseudospin pattern, $\cos{(k_rz+ \delta)}$, associated with pseudospins $\chi_{c1}$ and $\chi_{c2}$, but leaves  each DW's $\delta$ free to be determined through the  nonlocal cross-term interaction   $J_{12}$. The relative pseudospin alignment of $\chi_{c1}$ versus $\chi_{c2}$ is then set by each DW's choice of $\delta$.  This interaction is positive in the region between $r_2=0$ and  $r_2\approx\sqrt{2 \pi}w_0/4$,  and so the two pseudospins align such that $\sgn \{\chi_{c1} \chi_{c2}\}=+1$.  As the cross-term interaction strength approaches 0 near $r_2\approx\sqrt{2 \pi}w_0/4$, the relative phase between the DWs becomes uncorrelated and randomly fluctuates between 0 and $\pi$, reflecting the restoration of the original $\mathbb{Z}_{2}$ DW pattern symmetry. This can be seen by comparing the plot of $J_{12}$ in Fig.~\ref{fig4}g with the data.  For larger $r_2$'s,   $J_{12}$ changes sign, causing an antiferromagnetic alignment: $\sgn \{\chi_{c1} \chi_{c2} \}=-1$. This is manifest in a $\pi$ relative phase change between the light emitted from the two BECs~\footnote{We cannot determine the value of $\chi$, only the relative phase shift.}. To track this interaction sign change, we  measure the field phase at each  $r_2$ and plot the phase difference between the two sets of spots in  Fig.~\ref{fig4}f. 

We have demonstrated that the nonlocal interaction arising from Gouy phase anomalies in a confocal cavity offers a new tool to engineer cavity-mediated atom-atom interactions. Freezing the atoms into position, e.g., with an optical lattice, and coupling the atomic spins as in Ref.~\cite{Kroeze:2018wd}, would allow $U_\text{nonlocal}$ to mediate sign-changing spin-spin interactions of the form $\cos{(2\mbf{r}\cdot\mbf{r}^\prime/w_0^2)}$. 
This demonstration of sign-changing photon-mediated interactions, in conjunction with our recent demonstrations of spin-spin interactions~\cite{Kroeze:2018wd} and tunable-range atom-atom interactions~\cite{Vaidya:2018fp}---all within the same experimental apparatus---open the door to creating artificial spin glasses. With optical tweezers to place atoms in reproducible configurations~\cite{Endres:2016fk,Barredo:2016ea}, the exploration of replica symmetry breaking might be possible~\cite{FisherHertz}.  While replica symmetry breaking should be manifest in infinite-range spin glasses, the microscopic state of short-range spin glasses remains an outstanding question in statistical mechanics~\cite{stein2013spin}.  Moreover, placing atoms in specific locations to realize a particular graph of $\pm J_{ij}$ connectivity may provide a means for performing combinatorial optimization and Hopfield associative memory~\cite{Gopalakrishnan2011,Gopalakrishnan:2012cf,Torggler:2017hw,Rotondo:2017vg} in a quantum-optical setting.

We acknowledge funding support from the Army Research Office, the National Science Foundation under Grant No.~CCF-1640075, and the Semiconductor Research Corporation under Grant No.~2016-EP-2693-C.  J.~K. acknowledges support from SU2P.


%

\clearpage
\pagebreak

\section{Supplemental Material: \\ Sign-changing photon-mediated atomic interactions in multimode cavity QED}

\subsection{Spectrum of a confocal cavity}

Within paraxial optics, the beam inside a Fabry-Perot cavity is  described by Hermite-Gaussian modes. A mode $\Phi_{Q,l,m}$ is labeled by one longitudinal index $Q$ and two transverse indices $l$ and $m$. These indices count the number of field nodes along their respective axes. For a symmetric two-mirror cavity of length $L$, with $R$ as the mirror radius of curvature, the frequency of a given mode is
\be
f_{Qlm}=\frac{c}{2L}\big[ Q + \frac{l+m+1}{\pi} \arccos{ g} \big],
\ee
where $c$ is the speed of light inside the cavity, $g = 1-L/R$ and $c/2L$ is the free spectral range of the cavity. The term proportional to $\arccos{g}$ captures the effect of additional Gouy phase shifts on higher-order transverse modes,
which involve terms proportional to $(l+m+1)\psi(z)$, where $\psi(z) = \mrm{arctan} (z/z_R)$ is the Gouy phase and  $z_R$ is the Rayleigh range $z_R = \pi w_0^2/\lambda$. 

In general, different transverse modes will be resonant at different frequencies; however, degenerate cavities with special geometries can support a family of transverse modes, each with distinct spatial profiles, at a single frequency. In particular,  a confocal cavity has $L=R$ and thus $g = 0$.  Therefore, all modes that satisfy  the condition
\be
Q + \frac{1}{2}(l + m + 1) = Q_0 + \frac{(\eta+1)}{2}
\label{rescondition}
\ee
will be resonant at the same frequency $c(2 Q_0+\eta+1)/4L$, where $Q_0$ is a positive integer and $\eta=0 (1)$ for even (odd) families. At every half free spectral range, the transverse mode content varies between all even modes $l+m~\mrm{mod}~2 = 0$ and all  odd  modes $l+m~\mrm{mod}~2 = 1$. Within a degenerate resonance, to satisfy Eq.~\eqref{rescondition}, different transverse modes must carry different longitudinal indices. This causes the longitudinal profile of sequential transverse modes within a degenerate resonance to cycle between $+\cos{k_r z}$, $-\sin{k_r z}$, $-\cos{k_r z}$, and $+\sin{k_r z}$, as described in Fig.~\ref{fig1}(a) of the main text.

\subsection{Experimental apparatus}

This work employs a $R=1$-cm radius-of-curvature confocal cavity of length $L=R$. The length of the multimode cavity is adjustable~\cite{Kollar2015}, though in this work we set $L=R$.  We trap within this cavity a BEC of $2.5(3) {\times} 10^5$ $\mathrm{Rb}^{87}$ atoms in the $|F=1,m_F=-1 \rangle$ state.  See Ref.~\cite{Kollar2015} for BEC preparation procedure and Fig.~\ref{fig1} for illustration of experiment. The BEC is confined in a crossed optical dipole trap (ODT) formed by a pair of $1064$-nm laser beams propagating along $\hat{x}$ and $\hat{y}$ with waists of $40$~$\mu$m in the $xy$-plane and $80$~$\mu$m along  $\hat{z}$. The resulting trap frequencies of $(\omega_x,\omega_y,\omega_z) = 2 \pi \times [189(2),134(1),90(1)]$~Hz create a compact BEC with Thomas-Fermi radii $(R_x, R_y, R_z) = [4.2(1), 5.8(3), 8.9(1)]$~$\mu$m that are significantly smaller than the $w_0 = 35$~$\mu$m waist of the TEM$_{0,0}$ cavity mode. Acousto-optic deflectors (AODs) placed in the path of each ODT control the intensity and location of the ODTs, allowing us to translate the BEC to any point in the $xy$-plane with an uncertainty of $0.9~\mu$m. In the experiments of Figs.~\ref{fig3} and~\ref{fig4}, we use dynamic trap shaping~\cite{Henderson09} to produce two smaller BECs of $2.0(3) {\times} 10^5$ atoms each, with a population imbalance uncertainty of ${<}10$\%.  The relative position of these BECs along $\hat{x}$ is controlled by the AOD.  

Both the local oscillator beam (used for holographic imaging of the cavity emission) and the transverse pump are derived from the same laser but pass through different acousto-optic modulators (AOMs) for intensity stabilization. To maintain the relative phase stability between the two beams, both AOMs are driven by signals from the same multichannel direct digital synthesizer.  This synthesizer is synced to a stable Rb frequency reference. Due to path length drift, the relative phase between the pump and the local oscillator is  stable only within the same experiment realization.

\subsection{Holographic imaging}

The employed holographic imaging  method is described in detail in Ref.~\cite{Kroeze:2018wd} and is similar to that reported in Ref.~\cite{Schine:2018ui}.  Briefly,  a portion of the pump field---serving as a local oscillator (LO)---is directed onto the same EMCCD camera onto which the  cavity emission is imaged. The cavity field $E_c(\mbf{r}) = |E_c(\mbf{r})|e^{i \phi_c(\mbf{r})}$ and the LO field $E_\text{LO}(\mbf{r}) = |E_\text{LO}(\mbf{r})|e^{i \phi_\text{LO}(\mbf{r})}$ interfere to form a spatial heterodyne image $I_h(\mbf{r})$.  The image's interference fringes are proportional to the phase and amplitude of the cavity field:
\be
I_{h}(\mbf{r}) \propto  |E_c(\mbf{r})E_\text{LO}(\mbf{r})| \cos \left[ \Delta \mbf{k} \cdot \mbf{r} + \Delta\phi(\mathbf{r}) \right],
\label{hologram}
\ee
where  the phase difference between the cavity and LO wavefronts is $\Delta\phi(\mathbf{r}) =\phi_c(\mathbf{r}) -\phi_\text{LO}(\mathbf{r}) $.   The amplitude and phase of the fringes produced  are a measure of  $|E_c(\mbf{r})|$ and $\phi_c(\mbf{r})$.

Demodulating this image at the fringe wavevector $\Delta \mbf{k}$ provides a holographic reconstruction of $|E_c(\mbf{r})|$ and $\phi_c(\mbf{r})$. Accurate extraction of these images requires the correction of LO intensity and phase variation.  To do so for the confocal cavity, we perform a least-squares fit to the cavity emission intensity pattern using the exact theory result from Ref.~\cite{Vaidya:2018fp}.  We extract the LO phase variation from the difference between measured phase and the expected phase.

\subsection{Effective Hamiltonian}
The Green's function for the cavity-mediated interaction in a perfect confocal cavity near an even degenerate resonance can be written as a sum of the contributions from the two classes of longitudinal modes~\cite{Vaidya:2018fp,GouyPRA2018}: 
\begin{align}
4\mathcal{D}^{+}(\mbf{x},\mbf{x}') = 4\mathcal{D}^{+} (\mbf{r},\mbf{r^\prime},z,z^\prime) &=D_c (\mbf{r},\mbf{r^\prime})\cos{k_r z} \cos{k_r z^\prime} \nonumber \\ 
&+ D_{s} (\mbf{r},\mbf{r^\prime}) \sin{k_r z} \sin{k_r z^\prime},
\end{align}
with
\be
\begin{cases}
D_{c} =  \delta\Big(\frac{\sqrt{2}(\mbf{r} - \mbf{r^\prime})}{w_0}\Big) + \delta\Big(\frac{\sqrt{2}(\mbf{r} + \mbf{r^\prime})}{w_0}\Big) + \frac{1}{\pi} \cos\big(\frac{2 \mbf{r} \cdot \mbf{r^\prime}}{w^2_0}\big)    \\
D_{s} =  \delta\Big(\frac{\sqrt{2}(\mbf{r} - \mbf{r^\prime})}{w_0}\Big) + \delta\Big(\frac{\sqrt{2}(\mbf{r} + \mbf{r^\prime})}{w_0}\Big) - \frac{1}{\pi} \cos\big(\frac{2 \mbf{r} \cdot \mbf{r^\prime}}{w^2_0}\big). 
\end{cases}
\ee
To allow for the full phase freedom in the atomic density wave, the atomic profile is expanded as
\bea
\Psi(\mbf{x}) &&=\sqrt{\rho(\mbf{r})}\times \\\nonumber
&&\big[ \psi_0 +\sqrt{2}\cos{k_rx}( \psi_c \cos{k_rz} + \psi_s \sin{k_rz}) \big],
\eea
where for simplicity we shall assume a $\delta$-function transverse atomic profile $\rho(\mbf{r}) = \delta(\mbf{r} - \mbf{r}_0)$ since the Thomas-Fermi radius of the BEC is much smaller than the cavity waist $w_0$, $\mbf{r}_0$ is the location of the atoms in the cavity transverse plane, $\psi_0$ is the ground state fraction of the gas that has a uniform density profile (compared to the $\lambda$-scale) and $\psi_{c (s)}$ is the excited atomic density wave in the $\cos{k_rz}~(\sin{k_rz})$ pattern. The Hamiltonian is then 
\begin{align}\label{Ham}
H= E_0 \int &d^3\mbf{x} d^3\mbf{x^\prime} \cos(k_r x) \cos(k_r x') \times \nonumber \\
&|\Psi(\mbf{x})|^2 \mathcal{D}^{+}(\mbf{x},\mbf{x^\prime}) |\Psi(\mbf{x^\prime})|^2  \equiv -E_0 \mathcal{H}, 
\end{align}
where $E_0$ is a positive constant prefactor, and $\cos(k_r x) \cos(k_r x')$ term is due to the standing wave pump. Focusing only on the terms involving $\cos(2 \mbf{r} \cdot \mbf{r^\prime}/w^2_0)$ in $\mathcal{D}^{+}(\mbf{x},\mbf{x}')$, the effective Hamiltonian $\mathcal{H}$ can then be evaluated as
\be
\mathcal{H} = -\frac{1}{8 \pi}\left[ |\psi_0 \psi^{*}_{c} + \psi^{*}_0 \psi_{c}|^2 - |\psi_0 \psi^{*}_{s} + \psi^{*}_0 \psi_{s}|^2 \right] \cos\left(\frac{2 r^2_0}{w_0}\right).
\ee
Defining the following order parameters
\begin{align}
\chi_c &= \frac{\psi_0 \psi^{*}_{c} + \psi^{*}_0 \psi_{c}}{N} \nonumber \\
\chi_s &= \frac{\psi_0 \psi^{*}_{s} + \psi^{*}_0 \psi_{s}}{N},
\end{align}
and ignoring the numeric prefactor, we recover the effective Hamiltonian in the main text, where $N$ is the total atom number. For two BECs, the cross term in the integral in Eq.~\ref{Ham} gives rise to the interaction term
\be
H_{12} \propto -J_{12} (\chi_{c1} \chi_{c2} - \chi_{s1} \chi_{s2}),
\ee
where $J_{12} = 2N\cos\left({2\mbf{r_1} \cdot \mbf{r_2}}/{w^2_0}\right)$.

\end{document}